\documentclass[journal,10pt]{IEEEtran}
\IEEEoverridecommandlockouts
\usepackage[T1]{fontenc}
\usepackage{cite}
\usepackage{amssymb}
\usepackage{amsfonts}
\usepackage{fancyhdr}  %
\usepackage{extarrows}
\usepackage{algorithm}
\usepackage{multirow,tabularx}
\usepackage{mathtools}
\usepackage{xcolor}
\usepackage{bm}
\usepackage{algorithmic}
\usepackage{amsmath}
\usepackage{subfigure}
\usepackage{makecell}
\usepackage{colortbl}
\usepackage{eqparbox}

\ifCLASSINFOpdf
  \usepackage[pdftex]{graphicx}
  \graphicspath{{../pdf/}{../jpeg/}}
  \DeclareGraphicsExtensions{.pdf,.jpeg,.png}
\else
  \usepackage[dvips]{graphicx}
  \graphicspath{{../eps/}}
  \DeclareGraphicsExtensions{.eps}
\fi
\allowdisplaybreaks

\begin{document}

\title{Integrated Sensing, Communication, and Positioning in Cellular Vehicular Networks}

\author{Xin~Tong,~\IEEEmembership{Student Member,~IEEE, }
        Zhaoyang~Zhang,~\IEEEmembership{Senior Member,~IEEE, }
        Yuzhi~Yang,~\IEEEmembership{Member,~IEEE, }
        Yu~Ge,~\IEEEmembership{Member,~IEEE, }
        Zhaohui~Yang,~\IEEEmembership{Member,~IEEE, }
        Henk~Wymeersch,~\IEEEmembership{Fellow,~IEEE, }
        and~M\'{e}rouane~Debbah,~\IEEEmembership{Fellow,~IEEE}

\thanks{Copyright (c) 20xx IEEE. Personal use of this material is permitted. However, permission to use this material for any other purposes must be obtained from the IEEE by sending a request to pubs-permissions@ieee.org.}

\thanks{Xin~Tong, Zhaoyang~Zhang (Corresponding Author), and Zhaohui~Yang are with the College of Information Science and Electronic Engineering, Zhejiang University, Hangzhou 310027, China, and Zhejiang Provincial Key Laboratory of Multi-modal Communication Networks and Intelligent Information Processing, Hangzhou 310007, China. (e-mails: \{tongx, ning\_ming, yang\_zhaohui\}@zju.edu.cn)}

\thanks{Yuzhi~Yang and M\'{e}rouane~Debbah are with the College of Computing and Mathematical Sciences, Khalifa University, Abu Dhabi, UAE. (emails: \{yuzhi.yang, merouane.debbah\}@ku.ac.ae) }

\thanks{Yu~Ge and Henk~Wymeersch are with the Department of Electrical Engineering, Chalmers University of Technology, Gothenburg, Sweden. (emails: \{yuge, henkw\}@chalmers.se)}

\thanks{This work was supported in part by National Natural Science Foundation of China under Grants 62394292 and U20A20158, Zhejiang Provincial Key R\&D Program under Grant 2023C01021, Ministry of Industry and Information Technology under Grant TC220H07E, the Fundamental Research Funds for the Central Universities under Grant 226-2024-00069, and the SNS JU project 6G-DISAC under the EU's Horizon Europe research and innovation Program under Grant Agreement No 101139130.
}
}

\maketitle

\begin{abstract}

  In this correspondence, a novel integrated sensing and communication (ISAC) framework is proposed to accomplish data communication, vehicle positioning, and environment sensing simultaneously in a cellular vehicular network. By incorporating the vehicle positioning problem with the existing computational-imaging-based ISAC models, we formulate a special integrated sensing, communication, and positioning problem in which the unknowns are highly coupled. To mitigate the rank deficiency and make it solvable, we discretize the region of interest (ROI) into sensing and positioning pixels respectively, and exploit both the line-of-sight and non-line-of-sight propagation of the vehicles' uplink access signals. The resultant problem is shown to be a polynomial bilinear compressed sensing (CS) reconstruction problem, which is then solved by the alternating optimization (AO) algorithm to iteratively achieve symbol detection, vehicle positioning and environment sensing. Performance analysis and numerical results demonstrate the effectiveness of the proposed method.
\end{abstract}


\IEEEpeerreviewmaketitle
\section{Introduction}\label{js}
\IEEEPARstart{I}{n} the future sixth-generation (6G) wireless network, integrated sensing and communication (ISAC) \cite{Liuf} technology aims to use ubiquitous wireless communication signals to sense the environment information, including locations, shapes, and electromagnetic characteristics of targets in the environment. With emerging vehicle technologies, such as collision warning, autonomous driving, and intelligent traffic control \cite{Quf}, accurate environmental information is of vital importance. Moreover, the growing use of vehicle wireless signals presents opportunities for ISAC, making environment sensing and vehicle positioning in cellular vehicular networks become a key application \cite{chenx}.

Many research works have studied ISAC in vehicle communication scenarios. To list a few, \cite{zhangq} overcame the bottleneck of unreliable environment sensing caused by sensor failure and obstacle blockage and achieved high-performance vehicle cooperative communication and target detection. \cite{wangz} introduced a predictive beamforming scheme for vehicle tracking using vehicle identities and sensing echoes, enhancing communication performance.
\cite{gey} proposed a channel estimation and positioning algorithm that considers multipath interference, path merging, and round-trip-time (RTT) protocol and achieved sub-meter-accuracy vehicle positioning. Based on the known 3-dimensional (3D) environment map, \cite{bis} proposed a joint positioning and communication scheme in urban scenarios, which effectively solved the occlusion problem based on the LOS path formed by unmanned aerial vehicle assistance.

Recently, computational imaging \cite{Sun} as a potential environment sensing technology has recently been applied to wireless communication scenarios.
It obtains environmental information through pixel division, that is, dividing the unknown environment into discrete pixels as the smallest imaging unit and then calculating the value of each pixel. 
Compared with the existing environmental sensing methods in ISAC systems, computational imaging has the advantage of enabling imaging with high resolution, especially for tiny targets at the wavelength level, and thus is suitable for applications such as medical examinations and security checks \cite{Sun}.
Based on computational imaging and multiple access technology, a millimeter-wave ISAC system exploiting the sparsity of the environment is proposed in \cite{tongx}. The complex propagation characteristics of wireless signals, occlusion, and diffraction characteristics are jointly considered to achieve accurate imaging in wireless communication systems \cite{tongx2}. However, these works are based on the assumption of nearly perfect positioning, which limits their application in practical scenarios.

Different from existing works, in this paper, we aim to accomplish three different functionalities, including \textit{data communication, vehicle positioning, and environment sensing,} simultaneously in a cellular vehicular network. In particular, we incorporate the vehicle positioning problem with the existing computational-imaging-based ISAC models, which results in a special joint positioning and computational-imaging-based ISAC problem. By jointly processing the received signals of multiple base stations (BSs), a polynomial bilinear compressed sensing (CS) \cite{Donoho} reconstruction equation is solved by the proposed alternating optimization (AO) algorithm leveraging the relatively stronger power of the line-of-sight (LOS) path, and symbol detection, vehicle positioning, and environment sensing are jointly obtained.
The main contributions of this paper are summarized as follows: (i) We incorporate the vehicle positioning problem into a computational-imaging-based ISAC framework by discretizing the environment into positioning and sensing pixels, and propose a multiple access model that includes both vehicle position and environment information. (ii) We formulate the integrated sensing, communication, and positioning task as a polynomial bilinear CS reconstruction problem and propose an alternating optimization (AO) algorithm to iteratively solve for communication symbols, vehicle positions, and environmental information. (iii) We analyze the performance and the convergence of the proposed AO algorithm. Extensive numerical results verify the effectiveness of the proposed method.



\begin{figure}[t]
  \centering
  \includegraphics[width=0.48\textwidth]{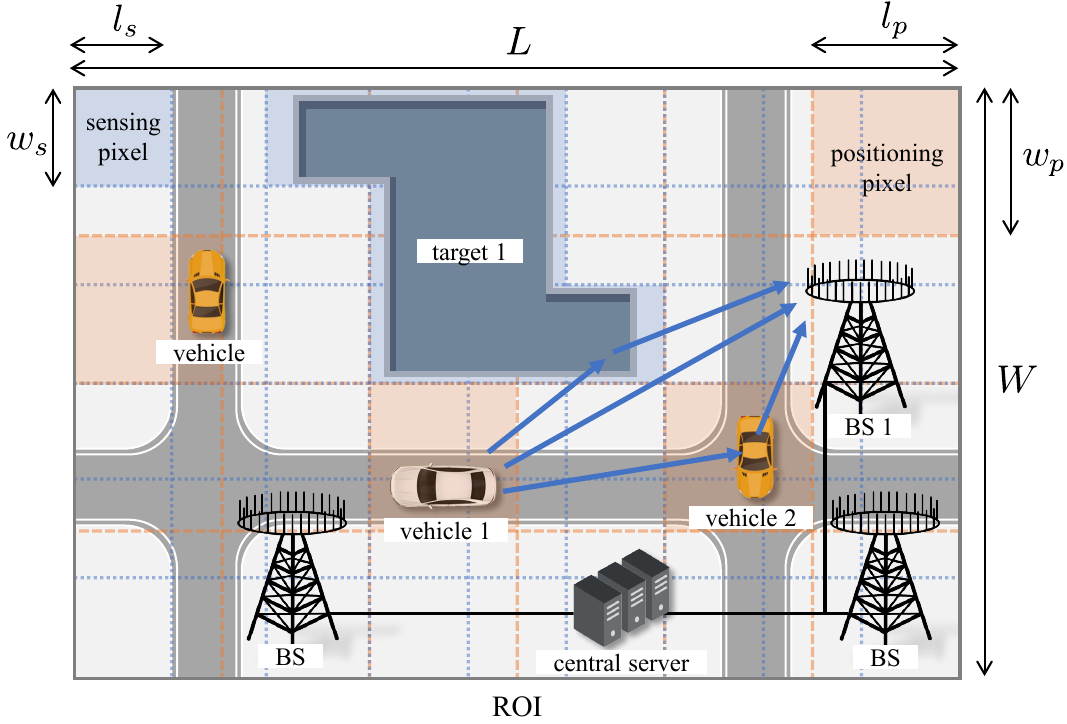}
  \caption{The uplink cellular vehicular wireless communication scenario.}
  \label{fig1}
  \end{figure}

\section{System Setting and Model}
\subsection{Environment Setting}\label{sec:ES}
We consider a 2-dimensional (2D)\footnote{The analysis and method for 3D scenarios can be derived using a similar approach.} wireless communication scenario, as depicted in Fig. \ref{fig1}, where joint vehicle positioning and ISAC are performed in the cellular vehicular networks. In the considered scenario, vehicle 1 sends uplink signals to BS 1. In addition to LOS propagation, compared with users such as mobile phones in conventional wireless communication scenarios, vehicles not only transmit signals but also significantly influence signal propagation. There are two types of non-line-of-sight (NLOS) propagation of communication signals\footnote{Higher-order interactions are also ignored and treated as noise due to the large attenuation of the mmWave NLoS path \cite{niu}.}: (a) vehicle-vehicle-BS, where the signal is scattered by other vehicles (vehicle 2) and propagated to BS 1, and vehicles will scatter each other's transmission signals; (b) vehicle-target-BS, where the signal is scattered by targets (target 1) in the environment and propagated to BS 1.
Based on the properties of LOS and NLOS propagation of signals in cellular vehicular scenarios, by jointly processing communication signals of multiple vehicles and multiple BSs in a central server, we aim to achieve vehicle positioning, communication symbol detection, and environment sensing jointly.

As shown in Fig. \ref{fig1}, the environment to be sensed is defined as the region of interest (ROI), which contains vehicles, BSs, and targets.
The ROI is evenly divided into two types of coexisting pixels, where vehicles are distributed in $N_{\rm p}$ orange positioning pixels (positioning reference units, PRUs), and targets are distributed in $N_{\rm s}$ blue sensing pixels. 
The length and width of each positioning pixel are $l_{\rm{p}}$ and $w_{\rm{p}}$, respectively, and the length and width of each sensing pixel are $l_{\rm{s}}$ and $w_{\rm{s}}$, respectively.
Let $L$ and $W$ denote the length and width of the ROI, respectively, thus, the total number of two types of pixels in the ROI is given by $N_{\rm{p}} =\frac{L}{l_{\rm{p}} }\cdot \frac{W} {w_{ \rm{P}}}$ and $N_{\rm{s}} =\frac{L}{l_{\rm{s}} }\cdot \frac{W} {w_{ \rm{s}}}$. It is assumed that the side length of the ROI is an integer multiple of the pixel side length.

For $N_{\rm{p}}$ positioning pixels, ${\bm p} = [p_1, p_2, \ldots, p_{N_{\rm{p}}}]^{\mathsf T} \in \{0, 1\}^{N_{\rm p} \times 1}$ indicates the presence of a vehicle at the corresponding pixel position. 
If there is a vehicle in the $n_{\rm{p}}$-th pixel, then $p_{n_{\rm{p}}} = 1$; otherwise, $p_{n_{\rm{p}}} = 0$. By solving for ${\bm p}$, We can locate vehicles without knowing the exact identities of the vehicles.
For the $n_{\rm{s}}$-th sensing pixel, it has the property of scattering coefficient $x_{n_{\rm{s}}}$.
If no target object is present in the $n_{\rm{s}}$-th pixel, then $x_{n_{\rm{s}}} = 0$; otherwise, $x_{n_{\rm{s}}} \in (0,1]$.
Therefore, the environmental information for the ROI can be characterized by the scattering coefficient vector, denoted as ${\bm{x}} = [x_1, x_2, \ldots, x_{N_{\rm{s}}}]^{\mathsf T} \in \mathbb{R}^{{N_{\rm{s}}} \times 1}$.

\subsection{System Model}\label{2B}
We consider the synchronization between the transceivers at the phase level and
adopt a specific communication modulation method, such as quadrature phase shift keying (QPSK) and quadrature amplitude modulation (QAM).
In a time slot, the uplink transmission symbols of multiple vehicles are ${\bm s} \in \mathbb{C}^{N_{\rm p} \times 1}$, where the power of ${\bm s} \in \mathbb{C}^{N_{\rm p} \times 1}$ is normalized. When there is a vehicle in the $n_{\rm p}$-th pixel, $s_{{n_{\rm p}}}$ represents the transmission symbol of the vehicle in this time slot, otherwise $s_{{n_{\rm p}}} = 0$. Obviously, the nonzero elements in ${\bm p}$ and ${\bm s}$ have the same index, that is, when ${s}_{n_{\rm p}} \neq 0$, ${p}_{n_{\rm p}} = 1$.

After propagation through the three types of channels in Section \ref{sec:ES}, the received symbols of $K$ BSs are denoted as ${\bm y} \in \mathbb{C}^{K \times 1}$. For the $k$-th BS, the received symbol ${y}_{k}$ of the current time slot is expressed as an unsourced multiple access model\footnote{Before this, some known sparse pilots should be used for synchronization between BSs and vehicles. There are some potential technologies \cite{vukmirovic} that motivate phase synchronization abilities.},
\begin{align}
    {y}_{k} & = {\bm p}^{\mathsf{T}}{\bf H}^{\rm LOS}_{k}{\bm s} + {\bm p}^{\mathsf{T}}{\bf H}^{\rm NLOS}_{k}{\bm s} + {\bm x}^{\mathsf{T}}{\bf H}^{\rm s}_{k}{\bm s} + n, \\
    & = {\bm p}^{\mathsf{T}}{\bf H}^{\rm V}_{k}{\bm s} + {\bm x}^{\mathsf{T}}{\bf H}^{\rm s}_{k}{\bm s} + n,\label{eq2}
\end{align}
where ${\bf H}^{\rm V}_{k} \in \mathbb{C}^{N_{\rm p} \times N_{\rm p}}$ is the vehicle communication multipath propagation gain from $N_{\rm p}$ positioning pixels through $N_{\rm p}$ positioning pixels to the $k$-th BS\footnote{We make the assumption that the transmission interval is sufficiently short to neglect any Doppler-induced phase effects.}. ${\bf H}^{\rm V}_{k}(i,j)$ represents the reference propagation gain of the signal transmitted by the vehicle in the $i$-th pixel after being scattered by the vehicle in the $j$-th pixel. Note that whether the vehicle actually exists in pixels $i$ and $j$ is determined by $\bm p$.
${\bf H}^{\rm V}_{k} \triangleq {\bf H}^{\rm LOS}_{k} + {\bf H}^{\rm NLOS}_{k} $, where ${\bf H}^{\rm LOS}_{k}$ and ${\bf H}^{\rm NLOS}_{k}$ represent the LOS and NLOS parts of the channel gain ${\bf H}^{\rm V}_{k}$, respectively. By definition, the diagonal elements of ${\bf H}^{\rm V}_{k}$ are from ${\bf H}^{\rm LOS}_{k}$, and the off-diagonal elements are from ${\bf H}^{\rm NLOS}_{k}$. This is because the diagonal elements represent the same pixels for the transmission and scattering.
${\bf H}^{\rm s}_{k} \in \mathbb{C}^{N_{\rm s} \times N_{\rm p}}$ is the multipath propagation gain from $N_{\rm p}$ positioning pixels through $N_{\rm s}$ sensing pixels to the $k$-th BS. ${\bf H}^{\rm s}_{k}(i,j)$ represents the reference propagation gain of the signal transmitted by the vehicle in the $i$-th pixel after being scattered by the scatterer in the $j$-th pixel. Note that whether the scatterer actually exists in pixels $j$ is determined by $\bm x$.
Gaussian white noise is represented by $n$. 

As shown in (2), we consider vehicles' position ${\bm p}$, communication symbols ${\bm s}$, and environment information ${\bm x}$ as unknown variables to be solved, and the other reference propagation gains are known values. The unknown variables ${\bm p}$ and ${\bm s}$ have the same position index, and ${\bm s}$ is independent of ${\bm x}$. It can be seen that in (2), ${\bm p}$ is bilinearly coupled with ${\bm s}$ and ${\bm x}$, which makes it difficult to separate and solve the unknown variables. We calculate ${\bf H}^{\rm V}_{k}$ and ${\bf H}^{\rm s}_{k}$ based on the conventional free space propagation model\footnote{We consider that the reference channel without the unknown environmental information can be modeled as a deterministic geometric channel. The free space propagation loss of the LOS path includes amplitude and phase. The NLOS propagation loss is obtained by multiplying multiple LOS parts \cite{tongx}.}. In the idea of dividing pixels for positioning and sensing, pixel scalability is a key factor in reducing computational complexity, thereby satisfying the hardware and latency requirements of vehicular applications. The proposed joint positioning and ISAC method is compatible with advanced techniques, such as the integral-form computational imaging model in \cite{tongicc}, which achieves phase error cancellation and extends computational imaging to large-scale wireless communication scenarios.
In addition, some novel propagation models have been proposed to more accurately describe wireless signal propagation in pixelated environments\cite{tongx, tongx2, xing}. In this paper, we focus on the joint performance of vehicle positioning and ISAC. For the propagation model, we can apply other accurate models and also consider other physical properties of pixels (e.g., shape of vehicle, frequency dependence) in the future.

\section{Problem Formulations and Algorithms}
\subsection{Problem Formulation}
In this section, received signals of multiple BSs are processed jointly to achieve vehicle positioning, communication symbol detection, and environment sensing. We formulate the optimization problem as
\begin{subequations}\label{eq3}
    \begin{align}
        \min_{{\bm x}, {\bm s}} \quad & \| {\bm x} \|_1 + \| {\bm s} \|_1 \\
        {\rm s.t.} \quad & \|{\bm y} - {\bf H}{\bm s}\|_2^2 \leq \varepsilon, \\
        &{p}_{n_{\rm p}} = 1\; {\rm when} \;{s}_{n_{\rm p}} \neq 0, \\
        & {\bm h}^{\mathsf{T}}_k = {\bm p}^{\mathsf{T}}{\bf H}^{\rm V}_{k} + {\bm x}^{\mathsf{T}}{\bf H}^{\rm s}_{k},
    \end{align}
\end{subequations}
where $n_{\rm p} \in \{1, \ldots, N_{\rm p}\}$, ${\bf H} = [{\bm h}_1; \ldots; {\bm h}_{K}]$ and $\varepsilon$ is a slack variable.

As shown in \eqref{eq2} and \eqref{eq3}, different from the equation ``${\bm y}={\bf A}{\bm x}$'' in the conventional CS problem\cite{Rangan, Vila, tongx2}, our problem equation ``${y}_k = {\bm x}_1^{\mathsf{T}} {\bf A}_k {\bm x}_2$'' is a \textit{bilinear} CS problem with two unknown sparse vectors. Currently, the general method for solving such a CS equation is theoretically challenging. Due to the non-convexity of the problem and the high coupling of unknown vectors, existing analytical methods are difficult to apply directly, and there are few algorithms designed for this type of problem \cite{Jiang, tianz}, as well as effective performance analysis to prove that the algorithm converges. We provide a practical application scenario for this difficult problem and explore potential solutions, which provides a basis for future algorithm design and theoretical research.

\subsection{Proposed Algorithm}
In this paper, we extend the generalized approximate message passing (GAMP)  algorithm \cite{Rangan, Vila} in the conventional CS problem to the proposed cellular vehicular communication scenario. Based on the prior that LOS propagation has greater power than NLOS propagation, the bilinear CS problem in \eqref{eq3} is solved by an AO method.

The conventional CS reconstruction problem can be solved by the GAMP algorithm using the iterative decomposition method, as shown in many related works\cite{tongx, tongx2}. Based on the measurements, the measurement matrix, and the prior information, the GAMP algorithm will give the maximum posterior probability (MAP) estimate of the unknown sparse variables, which will be performed in this section.

For the optimization problem in \eqref{eq3}, theoretically, the communication symbols are unknown discrete variables. By traversing the entire feasible domain, an upper bound on the performance can be achieved. However, the complexity of the traversal algorithm is impractical. As a second option, since the environmental information $\bm x$ and the communication symbols $\bm s$ are independent, they can be optimized separately. We consider the AO strategy and solve the environment information $\bm x$ and communication symbols $\bm s$ iteratively. In the $t$-th iteration, we decompose the optimization problem in \eqref{eq3} into the following two optimization sub-problems.

\begin{itemize}
    \item \textit{Vehicle positioning and symbol detection sub-problem:} Estimate communication symbols ${\bm{\hat s}}^{(t+1)}$ when the estimated environment information ${\bm{\hat x}}^{(t)}$ and the vehicle position ${\bm{\hat p}}^{(t)}$ is given in $t$-th iteration,
\begin{subequations}\label{op1}
    \begin{align}
        \min_{{\bm s}} \quad &  \| {\bm s} \|_1 \\
        {\rm s.t.} \quad & \|{\bm y} - {\bf H}{\bm s}\|_2^2 \leq \varepsilon, \\
        & {\bm h}^{\mathsf{T}}_k = ({\bm{\hat p}}^{(t)})^{\mathsf{T}}{\bf H}^{\rm V}_{k} + ({\bm{\hat x}}^{(t)})^{\mathsf{T}}{\bf H}^{\rm s}_{k},\\
        &{s}_{n_{\rm p}} = 0\; {\rm when} \;{p}^{(t)}_{n_{\rm p}} = 0.
    \end{align}
\end{subequations}

Since both vehicles and targets are sparsely distributed in the environment, 
the prior information of communication symbol $\bm s$ is modeled as follow a Bernoulli-Gaussian-mixture distribution, defined as
\begin{align}
  {p_{\rm p}}\left( {s_{n_{\rm p}}|{\bm{r}}_{\rm p}} \right) & = (1 - \gamma_{\rm p})\delta \left( s_{n_{\rm p}} \right) \nonumber \\ 
   + & \gamma_{\rm p}  \sum_{m = 1}^{M}{ {\bm \omega}(m){\cal N}\left( {s_{n_{\rm p}}|{\bm \theta}_{\rm p}(m) ,{{\bm \sigma}_{\rm p}(m)}} \right)},\label{eq5}
\end{align}
where ${\bm{r}_{\rm p}} \buildrel \Delta \over = \left[ {\gamma_{\rm p}, {\bm \omega}, {\bm \theta}_{\rm p}, {{\bm \sigma}_{\rm p}}} \right]$ denotes all parameters of the Bernoulli-Gaussian-Mixture distribution, $\gamma_{\rm p}$ is the sparsity coefficient of the vehicle distribution. There are $M$ distinct symbols, and ${\bm \omega}(m)$ is the proportion of the $m$-th symbol. In addition, ${\bm \theta}_{\rm p}(m)$ and ${{\bm \sigma}_{\rm p}(m)}$ represent the mean and the standard deviation of the $m$-th symbol, respectively. The parameters in \eqref{eq5} can be dynamically adjusted based on prior knowledge of road layout, directly affects weights during the iterative updates of the algorithm \cite{tongx}. The algorithm is not restricted to a specific scenario, but can flexibly accommodate different sources of prior information.

\item \textit{Sparse environment sensing sub-problem:} Given the communication symbols detection result ${\bm{\hat s}}^{(t+1)}$ and optimize the value of vehicle positioning results ${\bm{\hat p}}^{(t+1)}$ to be ${\hat p}^{(t+1)}_{n_{\rm p}} = 0$ when $|{s}^{(t+1)}_{n_{\rm p}}| \leq \beta^{(t)}$, where $\beta^{(t)}$ is a given threshold in $t$-th iteration. The environment sensing sub-problem with the estimated result ${\bm{\hat x}}^{(t+1)}$ is expressed as
\begin{subequations}\label{op2}
    \begin{align}
        \min_{{\bm x}} \quad & \| {\bm x} \|_1 \\
        {\rm s.t.} \quad & \|{\bm y} - {\bf H}{\bm{\hat s}}^{(t+1)}\|_2^2 \leq \varepsilon, \\
        & {\bm h}^{\mathsf{T}}_k = ({\bm{\hat p}}^{(t+1)})^{\mathsf{T}}{\bf H}^{\rm V}_{k} + {\bm x}^{\mathsf{T}}{\bf H}^{\rm s}_{k}.
    \end{align}
\end{subequations}

In addition, we model the prior environmental information $\bm x$ to follow a Bernoulli-Gaussian distribution in a limited interval\cite{tongx}, defined as
\begin{align}
  {p_{\rm s}}\left( {x_{n_{\rm s}}|{\bm{r}}_{\rm s}} \right) & = ( {1 - \gamma_{\rm s} + \lambda_{\rm s}} )\delta \left( x_{n_{\rm s}} \right) \nonumber \\ 
   + \gamma_{\rm s} & {\cal N}\left( {x_{n_{\rm s}}|\theta_{\rm s} ,{\sigma_{\rm s}}} \right)\big(u({x_{n_{\rm s}}}) - u({x_{n_{\rm s}}}-1)\big),\label{eq4}
\end{align}
where ${\bm{r}_{\rm s}} \buildrel \Delta \over = \left[ {\lambda_{\rm s}, \gamma_{\rm s}, \theta_{\rm s}, {\sigma_{\rm s}}} \right]$ denotes all parameters of the Bernoulli-Gaussian distribution, $\delta \left(  \cdot  \right)$ is the Dirac function, $u\left(  \cdot  \right)$ is the step function and $\lambda_{\rm s}  = \gamma_{\rm s}\int_{x \in \left( { - \infty ,0} \right] \cup \left[ {1, + \infty } \right)} {{\cal N}\left( {x|\theta_{\rm s} ,{\sigma_{\rm s}}} \right)} {\rm d}x$. The sparsity of the environment is denoted as $\gamma_{\rm s}$, and higher sparsity means more targets in the environment. In addition, $\theta_{\rm s}  \in \left[ {{\rm{0}},{\rm{1}}} \right]$ and ${\sigma_{\rm s}}$ represent the mean and the standard deviation of the environmental information $x_{n_{\rm s}}$, respectively. The parameters in \eqref{eq4} can be dynamically adjusted based on the the continuous layout of targets, as shown in our previous work \cite{xing, zhangy}.

\end{itemize}

\begin{algorithm}[h]
    \caption{The Proposed AO Algorithm}
    \label{alg1}
    \begin{algorithmic}[1]
    \REQUIRE
    Given the received symbols $\bm y$, theoretical propagation gains ${\bf H}^{\rm V}_{k}$ and ${\bf H}^{\rm s}_{k}$.

    \ENSURE
    Estimated sparse environment target ${\bm {\hat x}}^{(t)}$. The position and symbols of each vehicle ${\bm {\hat p}}^{(t)}$ and ${\bm {\hat s}}^{(t)}$.
    
    \STATE
    \textbf{Initialization}: Set prior parameter ${\bm r}_{\rm s}$ and ${\bm r}_{\rm p}$. Set $t = 0$, ${\bm{\hat p}}^{(0)}_{n_{\rm p}} = 1$, ${\hat x}^{(0)}_{n_{\rm s}} = 0$, ${\hat s}^{(0)}_{n_{\rm p}} = 0$.
    
    \WHILE {$\|{\bm y} - {\bf H}{\bm{\hat s}}^{(t)}\|_2^2 > \varepsilon $, where ${\bm h}^{\mathsf{T}}_k = ({\bm{\hat p}}^{(t)})^{\mathsf{T}}{\bf H}^{\rm V}_{k} + ({\bm{\hat x}}^{(t)})^{\mathsf{T}}{\bf H}^{\rm s}_{k}$,}

    \STATE
    \textit{Solve optimization problem \eqref{op1} by GAMP:}
    
    Input the prior probability in \eqref{eq5} and plug ${\bm{\hat p}}^{(t)}$ and ${\bm{\hat x}}^{(t)}$ into \eqref{op1} to estimate ${\bm{\hat s}}^{(t+1)}$.

    \STATE
    Set ${\hat p}^{(t+1)}_{n_{\rm p}} = 0$ when $|{s}^{(t+1)}_{n_{\rm p}}| \leq \beta^{(t)}$, where $\beta^{(t)}$ is a given threshold in $t$-th iteration.

    \STATE
    \textit{Solve optimization problem \eqref{op2} by GAMP:}
    
    Input the prior probability in \eqref{eq4} and plug ${\bm{\hat s}}^{(t+1)}$ and ${\bm{\hat p}}^{(t+1)}$ into \eqref{op2} to estimate ${\bm{\hat x}}^{(t+1)}$.
    
    \STATE
    ${t} = {t} + 1.$

    \ENDWHILE
    \end{algorithmic}
\end{algorithm}

In the proposed AO algorithm, we perform optimization problems \eqref{op1} and \eqref{op2} alternately, and the results of the two optimization problems promote each other and converge to accurate sensing results, which will be analyzed in Section \ref{sec:ana}. The specific AO algorithm is summarized in Algorithm 1. The computational complexity of the proposed algorithm is expressed as $\mathcal{O}(N_{\rm p}K + N_{\rm s}K)$, which includes the computational complexity of two sub-problems. During the algorithm execution, the number of pixels $N_{\rm p}$ and $N_{\rm s}$ should be appropriately selected to limit the computational complexity.

\section{Performance Analysis}\label{sec:ana}
Different from the quantization error of the pixelated model for the practical scenario in Section \ref{2B}, this section focuses on the estimation error of solving unknown variables. The sources of estimation error include limitations in system resources as well as various random factors.
According to the basic theory of CS, some related work \cite{tongx, tongx2} has analyzed the relationship between system resources and sensing performance, such as the number of antennas, signal-to-noise ratio (SNR), etc. 
Various random factors such as channel fading, vehicle mobility, and measurement noise, which are common in real-world cellular vehicular network environments. Among them, channel randomness and measurement errors can be modeled as noise terms in \eqref{eq2} due to their Gaussian random characteristics, and can be effectively addressed using the CS method. In contrast, the estimation error induced by vehicle mobility becomes the dominant random factor. Let the projection of the vehicle’s velocity in the angle-of-arrival direction be denoted as $v$, then the resulting Doppler shift is expressed as $e^{\jmath 2\pi T v/\lambda}$, where $T$ denotes the transmission duration of each symbol, and $\lambda$ denotes the carrier wavelength. 
In this work, we assume that $T$ is sufficiently small such that the resulting error is negligible. However, in practical scenarios where the displacement $vT$ becomes comparable to the carrier wavelength, this Doppler-induced error can significantly degrade system performance \cite{tongicc, xiao}. Addressing positioning problems involving velocity estimation is left for future work.
In this section, we focus on analyzing the performance and AO motivation of Algorithm 1. Specifically, we show how the estimation errors affect the convergence progress.

\subsection{Positioning and Communication Performance}\label{sec:pcp}

Denote the positioning error of ${\bm {\hat p}}^{(t)}$ as ${\bm q}^{(t)} \in \{0, \pm 1\}^{N_{\rm p} \times 1}$, i.e., ${\bm q}^{(t)} = {\bm {\hat p}}^{(t)} - {\bm p}$.
By analyzing the positioning error ${\bm q}^{(t)}$, we can show the improvement of positioning and communication performance during the iteration.
Therefore, in the $t$-th iteration, we reformulate $({\bm h}^{(t)}_{k})^{\mathsf{T}}{\bm s}$ in \eqref{op1} as
\begin{equation}
    ({\bm h}^{(t)}_{k})^{\mathsf{T}}{\bm s} = ({\bm p}+{\bm q}^{(t)})^{\mathsf{T}}{\bf H}^{\rm V}_{k}{\bm s} + ({\bm x^{(t)}})^{\mathsf{T}}{\bf H}^{\rm s}{\bm s} - ({\bm q}^{(t)})^{\mathsf{T}}{\bf H}_{k}^{\rm V}{\bm s}.\label{eq9}
\end{equation}
At the same time, we define a power ratio $\delta^{(t)}$ in the $t$-th iteration as
\begin{equation}
    \delta^{(t)} = \frac{\|({\bm p}+{\bm q}^{(t)})^{\mathsf{T}}{\bf H}_{k}^{\rm V}{\bm s}\|_2^2}{\|({\bm x^{(t)}})^{\mathsf{T}}{\bf H}_{k}^{\rm s}{\bm s} - ({\bm q}^{(t)})^{\mathsf{T}}{\bf H}_{k}^{\rm V}{\bm s}\|_2^2} = \frac{\| {\bm h}_{{\rm A},k}^{(t)}{\bm s} \|_2^2}{\| {\bm h}^{(t)}_{{\rm B},k}{\bm s} \|_2^2},
\end{equation}
where ${\bm h}_{{\rm A},k}^{(t)} = ({\bm p}+{\bm q}^{(t)})^{\mathsf{T}}{\bf H}_{k}^{\rm V}$ represents the known part of the model for solving \eqref{op1}, and ${\bm h}_{{\rm B},k}^{(t)} = ({\bm x^{(t)}})^{\mathsf{T}}{\bf H}_{k}^{\rm s} - ({\bm q}^{(t)})^{\mathsf{T}}{\bf H}_{k}^{\rm V}$ represents the unknown part caused by the environment and positioning errors. Therefore, $\delta^{(t)}$ affects the performance of estimating ${\bm{\hat s}}^{(t+1)}$ in the step 3 of Algorithm 1.

Here we analyze the starting conditions of the proposed AO algorithm. At the beginning of the iteration, ${\bm p}^{(0)}$ is set to be all ones. Therefore, there is no overlap between the support sets of ${\bm p}$ and ${\bm q}^{(0)}={\bm p}^{(0)}-{\bm p}$.
We can conclude that $ ({\bm q}^{(0)})^{\mathsf{T}}{\bf H}_{k}^{\rm V}{\bm s} = ({\bm q}^{(0)})^{\mathsf{T}}{\bf H}_{k}^{\rm NLOS}{\bm s}$.
Since the LOS propagation power is much greater than the NLOS propagation power related to vehicles and targets, we know $\|{\bm p}^{\mathsf{T}}{\bf H}_{k}^{\rm V}{\bm s}\|_2^2\gg\|({\bm q}^{(0)})^{\mathsf{T}}{\bf H}_{k}^{\rm V}{\bm s}\|_2^2$, thus $\delta^{(0)} \gg 1$.
The above analysis shows that ${\bm h}_{{\rm A},k}^{(0)}$ dominates the $\bm{s}$-related items in the received signal in the initial stage of iteration. Therefore, we can treat ${\bm h}_{{\rm B},k}^{(0)}$ as an error, and we can obtain an effective preliminary result ${\bm{\hat s}}^{(0)}$ from step 3 in Algorithm 1 based on ${\bm h}_{{\rm A},k}^{(0)}$. 

After the iteration starts, we further analyze the positioning and communication performance. The above analysis shows that false alarm (detect ${\hat p}^{(t)}_{n_{\rm p}} = 1$ when $p_{n_{\rm p}} = 0$) error does not have much impact on system performance because it hardly affects $\delta^{(t)}$. Since we use all-ones initialization and the target is always sparse, the missing detection (detect ${\hat p}^{(t)}_{n_{\rm p}} = 0$ when $p_{n_{\rm p}} = 1$) error is limited until the system has reached a near-optimal point.
Thus, with the iteration, the positioning error ${\bm q}^{(t)}$ gradually decreases. At the same time, in step 4 of Algorithm 1, We should not set $\beta^{(t)}$ too large so that $ ({\bm q}^{(t)})^{\mathsf{T}}{\bf H}_{k}^{\rm V}{\bm s} = ({\bm q}^{(t)})^{\mathsf{T}}{\bf H}_{k}^{\rm NLOS}{\bm s}$ holds, which can ensure that $\delta^{(t)}$ continues to increase and gradually improve the algorithm performance. In practical scenarios, $\beta^{(t)}$ should be set according to the convergence of the algorithm.

\subsection{Sensing Performance}
After obtaining the positioning and symbol detection results, the sparse environment sensing sub-problem \eqref{op2} is carried out. Let error ${\bm r}^{(t)} \in \{0, 1\}^{N_{\rm p} \times 1}$ represent the symbol detection error in the $t$-th iteration, that is, ${\bm {\hat s}}^{(t)} = {\bm s} + {\bm r}^{(t)}$. 
By analyzing the symbol detection error ${\bm r}^{(t)}$, we can show the improvement of sensing performance during the iteration.
We plug  ${\bm {\hat s}}^{(t)}$ into \eqref{eq9} and put the known terms on the left side of the equation to obtain the CS reconstruction equation for ${\bm x}^{(t)}$,
\begin{align}
    &({\bm h}^{(t)}_{k})^{\mathsf{T}}{\bm {\hat s}}^{(t)} - ({\bm {\hat p}}^{(t)})^{\mathsf{T}}{\bf H}^{\rm V}_{k}{\bm {\hat s}}^{(t)} =  ({\bm x^{(t)}})^{\mathsf{T}}{\bf H}^{\rm s}{\bm {\hat s}}^{(t)}  \nonumber \\
    &\;\;\;\;- ({\bm x^{(t)}})^{\mathsf{T}}{\bf H}^{\rm s}{\bm {r}}^{(t)}
    - ({\bm q}^{(t)})^{\mathsf{T}}{\bf H}_{k}^{\rm V}{\bm {\hat s}}^{(t)} - {\bm p}^{\mathsf{T}}{\bf H}_{k}^{\rm V}{\bm {r}}^{(t)}.\label{eq9}
\end{align}
The $({\bm x^{(t)}})^{\mathsf{T}}{\bf H}^{\rm s}{\bm {\hat s}}^{(t)}$ term on the right side of the equation represents the linear equations for solving ${\bm x}^{(t)}$, and the remaining three terms are noise term, representing three aspects that degrade sensing performance. The noise term $({\bm x^{(t)}})^{\mathsf{T}}{\bf H}^{\rm s}{\bm {r}}^{(t)}$ contains the variable ${\bm {r}}^{(t)}$, whose power depends on the accuracy of symbol detection. The noise term $({\bm q}^{(t)})^{\mathsf{T}}{\bf H}_{k}^{\rm V}{\bm {\hat s}}^{(t)}$ contains $({\bm q}^{(t)})^{\mathsf{T}}{\bf H}_{k}^{\rm V}$, whose power depends on the accuracy of the positioning and we make it small during iterations as described in Section \ref{sec:pcp}. The power of the noise term ${\bm p}^{\mathsf{T}}{\bf H}_{k}^{\rm V}{\bm {r}}^{(t)}$ is not only related to the symbol detection accuracy, but also to the power of the LOS path. We can conclude that positioning and communication benefit from the strong power of the LOS path, but it relatively drowns out the power of the NLOS path, which degrades the sensing performance.

\begin{figure*}[t]
  \centering
  \begin{minipage}[t]{1\textwidth}
      \centering
      \subfigure[The positioning accuracy.]{
      \includegraphics[height=4.4cm]{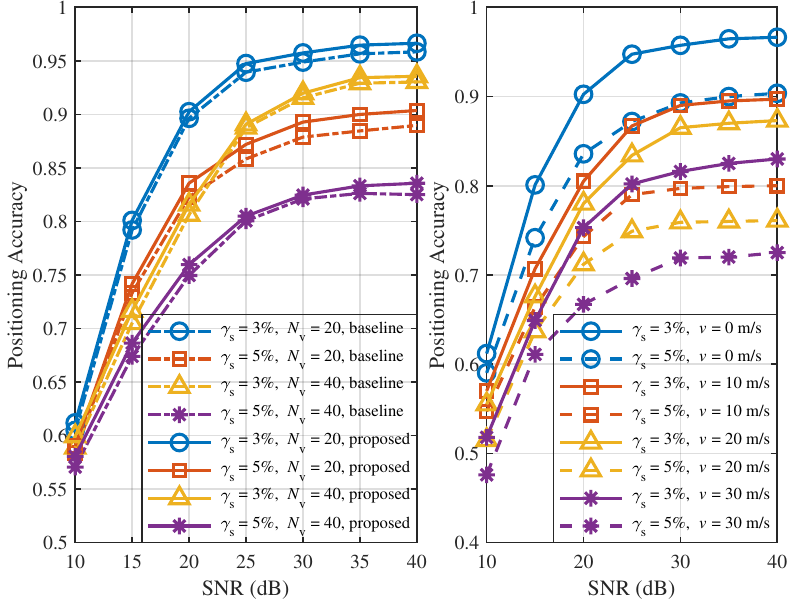}}
      \subfigure[The SER of symbol detection.]{
      \includegraphics[height=4.4cm]{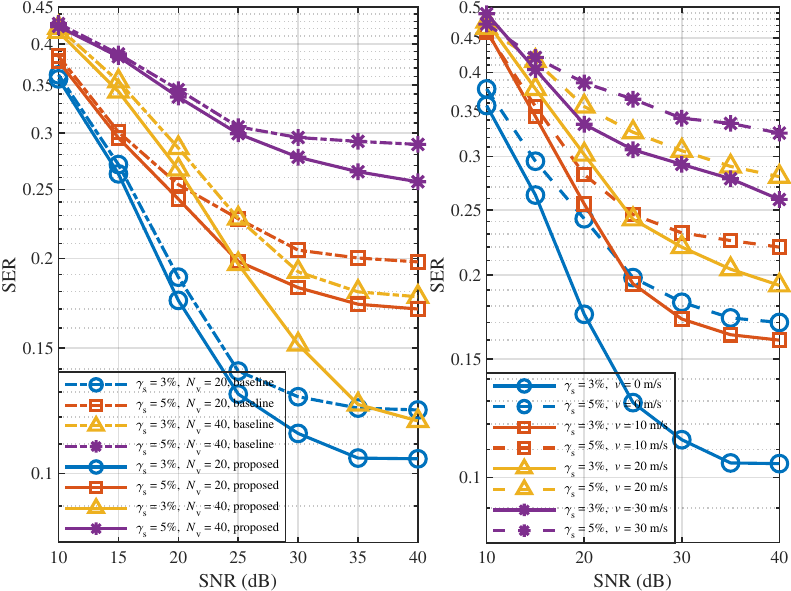}}
      \subfigure[The MSE of environment sensing.]{
      \includegraphics[height=4.4cm]{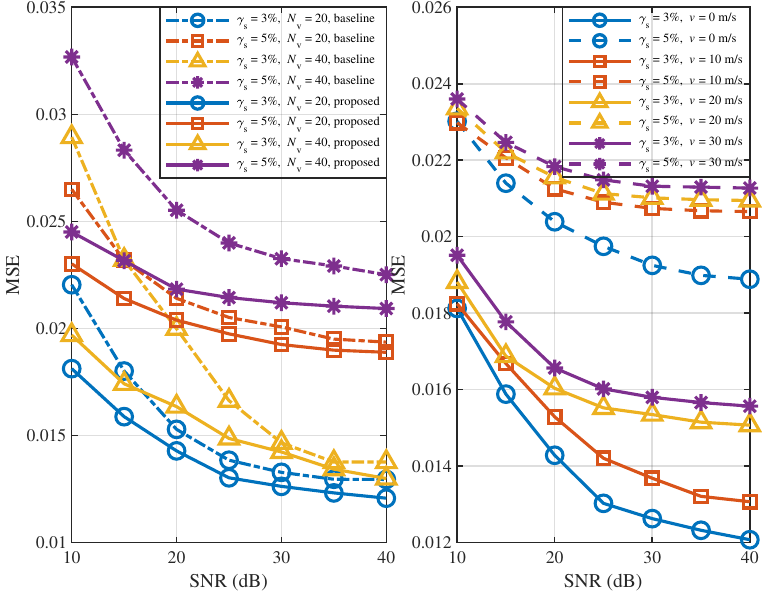}}
      \caption{The relationship between the SNR and the system performance for different sparsity levels $\gamma_{\rm s}$, different numbers of vehicles $N_{\rm v}$ and different average vehicle speeds $v$.}
      \label{fig2}
  \end{minipage}
\end{figure*}

\begin{figure*}[t]
  \centering
  \begin{minipage}[t]{1\textwidth}
      \centering
      \subfigure[The positioning accuracy.]{
      \includegraphics[height=4.6cm]{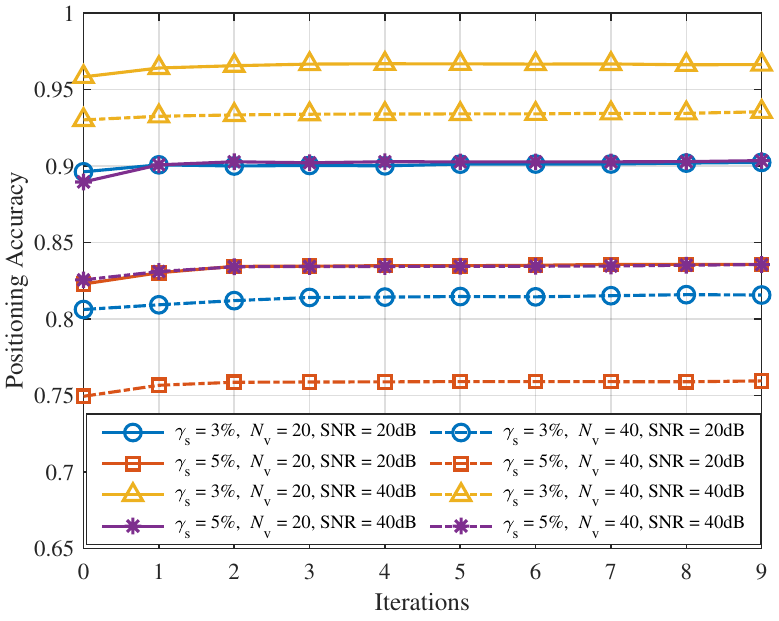}}
      \subfigure[The SER of symbol detection.]{
      \includegraphics[height=4.6cm]{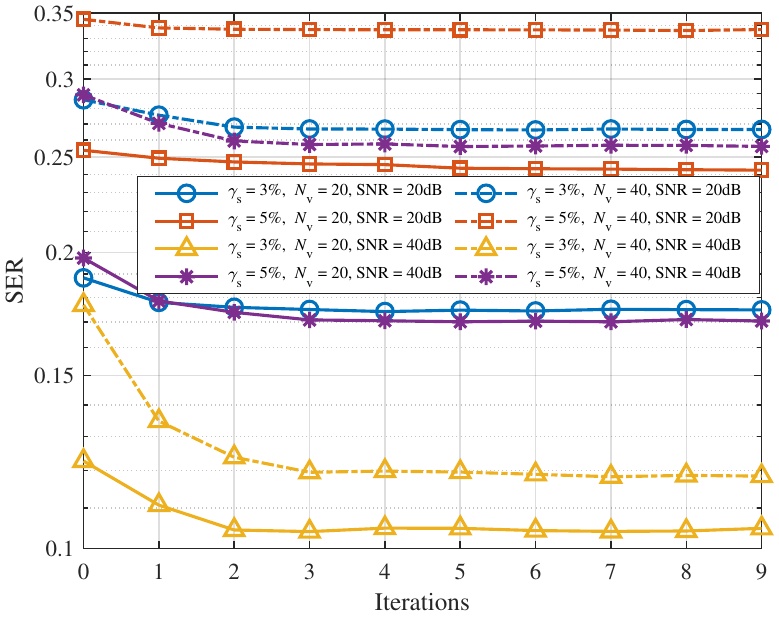}}
      \subfigure[The MSE of environment sensing.]{
      \includegraphics[height=4.6cm]{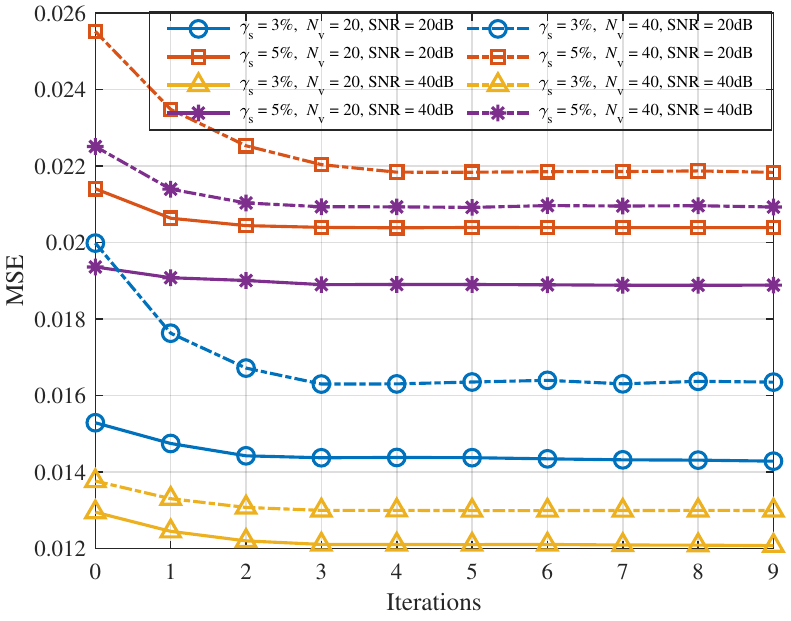}}
      \caption{Convergence of the proposed AO algorithm for different sparsity levels $\gamma_{\rm s}$, different SNR and different number of vehicles $N_{\rm v}$.}
      \label{fig3}
  \end{minipage}
\end{figure*}


\section{Numerical Results}
In this section, we present the numerical simulation results of the proposed method and provide a comparison with a state-of-the-art compressed sensing reconstruction method, GAMP \cite{Rangan, Vila}, which directly solves sub-problems \eqref{op1} and \eqref{op2} without AO as the baseline. In addition, the AO strategy will increase the amount of computation by $T$ times compared to the baseline, where $T$ is the number of iterations.

We consider a 2D scenario with a square ROI in which targets and vehicles are randomly distributed, as shown in Fig.~\ref{fig1}. We set $K = 50$, $N_{\rm s} = 15\times 15$ and $N_{\rm p} = 10\times 10$. Threshold $\beta^{(t)} \in [0.5, 0.9]$, and gradually increases with iterations. The sizes of pixels are set to $l_{\rm s} = w_{\rm s} = 1~{\rm m}$ and $l_{\rm p} = w_{\rm p} = 1.5~{\rm m}$. The carrier frequency is set to $\rm 30$~GHz. The scattering coefficient of the targets is set to 1, and the QPSK constellation is used as the communication symbol. Based on this, the prior amplitudes in ${\bm{r}_{\rm s}}$ and ${\bm{r}_{\rm p}}$ are set, and the standard deviations are estimated by the expectation maximization (EM) algorithm \cite{Vila}. The detection rate is used to measure the positioning accuracy and is expressed as ${\bm p}^{\mathsf{T}}{\hat {\bm p}} / N_{\rm p}$, which is the true positive rate of the detection. The symbol error rate (SER) between the original transmitted symbols and the detected symbols is used to measure the communication performance. The mean square error (MSE) between the original environment and the reconstructed environment is used to measure the sensing performance. 

Fig.~\ref{fig2} shows the relationship between the signal-to-noise ratio (SNR) and system performance. SNR is defined as the power ratio of the received symbol to the noise in \eqref{eq2}. The number of vehicles is denoted as $N_{\rm v}$. It can be concluded that as the SNR increases, the system performance improves. At the same time, as the sparsity $\gamma_{\rm s}$ of the environment increases, the system performance decreases. When SNR increases, the communication performance curve becomes flat because the positioning and sensing performance reach their limit. The simulation results also illustrate that the performance of the proposed AO algorithm is better than the baseline. 
Fig. \ref{fig2} also illustrates the impact of Doppler effects on system performance under average vehicle speeds of $v = \rm 10~m/s$, $\rm 20~m/s$, and $\rm 30~m/s$. It can be concluded that as the vehicle speed increases, Doppler weakens the system performance more significantly, but the proposed algorithm remains effective.
Fig. \ref{fig3} shows the convergence of the proposed AO algorithm. It can be concluded that, under different parameter settings, with the iteration of the proposed AO algorithm, the system performance gradually improves and converges from the baseline. Let the transmission duration of each symbol be $T = 1/30~{\rm kHz}$.

\section{Conclusion}
In this paper, we proposed a novel multiple access model that includes environmental and vehicle position information. By incorporating the vehicle positioning problem with the existing computational-imaging-based ISAC models, we simultaneously solve for data communication, vehicle positioning, and environment sensing via a polynomial bilinear CS equation. The proposed AO algorithm demonstrates effective performance in all three tasks.

\ifCLASSOPTIONcaptionsoff
  \newpage
\fi

\bibliographystyle{IEEEbib}
\bibliography{IEEEabrv, ref}
\end{document}